# Abrupt changes in the graphene on Ge(001) system at the onset of surface melting

L. Persichetti[1], L. Di Gaspare[1,*], F. Fabbri[2], A. M. Scaparro[1], A. Notargiacomo[3], A. Sgarlata[4], M. Fanfoni[4], V. Miseikis[2], C. Coletti[2], and M. De Seta[1]

[1]Dipartimento di Scienze, Università Roma Tre, Viale G. Marconi, 446- 00146 Roma, Italy

[2]Center for Nanotechnology Innovation @NEST, IIT, Piazza San Silvestro 12, 56127 Pisa, Italy

[4]Dipartimento di Fisica, Università di Roma "Tor Vergata", Via Della Ricerca Scientifica, 1- 00133 Roma, Italy

[3]Institute for Photonics and Nanotechnology, Via Cineto Romano 42, 00156, CNR-Rome, Italy

By combining scanning probe microscopy with Raman and x-ray photoelectron spectroscopies, we investigate the evolution of CVD-grown graphene/Ge(001) as a function of the deposition temperature in close proximity to the Ge melting point, highlighting an abrupt change of the graphene's quality, morphology, electronic properties and growth mode at 930 °C. We attribute this discontinuity to the incomplete surface melting of the Ge substrate and show how incomplete melting explains a variety of diverse and long-debated peculiar features of the graphene/Ge(001), including the characteristic nanostructuring of the Ge substrate induced by graphene overgrowth. We find that the quasi-liquid Ge layer formed close to 930 °C is fundamental to obtain high-quality graphene, while a temperature decrease of 10 degrees already results in a wrinkled and defective graphene film.

*corresponding author:

Fax: +39 0657333431 E mail: luciana.digaspare@uniroma3.it (Luciana Di Gaspare)

## 1. Introduction

Over the last three decades, revolutionary advances in technology have been paralleled and propelled by breakthroughs in silicon-based complementary-metal-oxide-semiconductor (CMOS) technology through a continued miniaturisation of silicon integrated circuits [1]. This positive trend is, however, nearing the intrinsic limits of chip scaling at the nanoscale, thus calling for alternative materials to promote beyond-CMOS evolutionary advances [2]. Rather than replacing silicon, the way to achieve this goal is to develop integration schemes which make novel materials complementary to conventional CMOS technologies. Graphene is a prime candidate for non-silicon electro-optical technology, having already demonstrated its potential for a vast range of applications, such as data communications [3], high-performance LEDs [4], low-power photonics [5, 6] and ultrafast- and broad-band photodetection from the ultraviolet to the terahertz range [7-9]. The main bottleneck to unlocking graphene's potential is developing metal-free deposition techniques compatible with the standard CMOS paradigm [10]. To this end, the catalysed chemical vapor deposition (CVD) of graphene on Ge or Ge/Si substrates [11-18] is a promising route for overcoming the issue of metallic contamination inherently associated with the well-established CVD synthesis of graphene on metals or with the transfer process from metallic templates [19-21]. Besides being intrinsically compatible with Si-based processing technology, Ge is well-suited as a catalytically-active substrate for promoting the decomposition of carbon precursors and the formation of a graphene phase, due to the low solubility of C in Ge and the absence of a stable Ge carbide. Despite these favorable premises, the exploitation of CVD-grown graphene/Ge for device-oriented applications is still hindered by a problematic understanding of the complex interplay between the quality and morphology of the graphene film and alterations in the surface structure morphology of Ge resulting from the graphene growth. On the technologically-attracting Ge(001) face, the most striking manifestation of this puzzling growth mechanism is the observation of a peculiar faceting pattern along {10L} crystal faces of Ge [14, 17, 22-24] which is selectively formed on the substrate only below the graphene flakes [25]. The presence of the faceting on Ge is intimately connected to the quality of the graphene layer in terms of electronic and transport properties [22, 26, 27]. Nonetheless, identifying the driving force for Ge nanofaceting and its correlation to the structural changes occurring in graphene at the atomic-scale is problematic and still remains a puzzling issue, due to the large number of parameters contributing to the growth process. The main hurdle is that slight variations of these parameters among the growth recipes used in literature lead to controversial results [11, 13-15, 17, 18, 22-24, 28], hampering a clear-cut picture. To disentangle the role of the growth temperature, in this study we investigate the evolution of graphene and, in parallel, of the Ge surface while the growth temperature is varied and controlled within a few tens of degrees from the melting point of Ge, through an accurate thermal control-loop feedback mechanism and a proper-designed multi-step temperature ramp in the CVD process. By combining atomic-force (AFM) and scanning tunneling microscopy (STM) with Raman, scanning electron microscopy (SEM) and X-ray photoelectron (XPS) spectroscopies, we find that, close to the Ge melting point, small variations in the deposition temperature of less than 30 degrees dramatically affect the quality of the graphene adlayer. Between 930 and 910 °C, we observe an evolution from a flat and almost defect-free graphene layer to a more wrinkled and defective graphene at lower temperatures. We correlate this abrupt structural transition in graphene to the incomplete surface melting behavior of the Ge substrate. In addition, we show that this incomplete melting of Ge is pivotal in explaining the characteristic nanofaceting of the Ge(001) surface developing underneath the growing graphene film.

## 2. Materials and methods

 The graphene films were grown on Ge(001) substrates (N-type Sb-doped, $n=10^{16}$ cm$^{-3}$) by employing a commercially available CVD reactor (Aixtron BM). Prior to graphene deposition, the Ge substrates were cleaned *ex-situ* by several rinsing and drying steps using isopropyl alcohol and de-ionized water. Graphene was grown at 100 mbar from a CH$_4$/H$_2$ gas mixture with Ar as carrier gas. The CH$_4$, H$_2$ and Ar fluxes were set at 2, 200, and 800 sccm, respectively. The deposition temperature $T_D$ was varied between 910 and 930 °C and the cooling down to room temperature was performed in $H_2$ and Ar only. Temperatures 5° C higher than 930 °C resulted in poorer graphene quality and a partial melting of the Ge substrate. Using the CVD parameters described above, at $T_D$= 930 °C and for a deposition time $t_D$= 60 min, we obtained a state-of-the-art single-layer graphene, grown in a layer-by-layer mode, which we will take as a reference sample [14]. Here we performed an accurate and fine reduction of the growth temperature to highlight its impact on the growth process. In order to accurately control the temperature close to the Ge melting point, the heating of the substrate was carried out using a multi-step temperature ramp with a rate progressively decreasing as $T_D$ is approached. Two graphite heaters ensure a reproducible and homogeneous temperature over the active growth region.

The XPS measurements were performed using a monochromatic Al $K_\alpha$ source ($hv$=1486.6 eV) and a concentric hemispherical analyzer operating in retarding mode (Physical Electronics Instruments PHI), with overall resolution of 0.4 eV. The carbon amount $\rho$ deposited on the sample was evaluated by using the $C_{1s}$ core level area intensity normalized to that acquired in the same experimental conditions on a commercial graphene monolayer (CGM) (i.e. CVD graphene grown on copper foil and transferred on $SiO_2$) mounted next to the analyzed sample, i.e. $\rho = I_{C_{1s}}^{sample} / I_{C_{1s}}^{CGM}$ .

Raman spectroscopy (Renishaw inVia confocal Raman microscope) was carried out using an excitation wavelength of 532 nm, a 100$x$ objective and a laser spot size of 3 µm. The peak parameters were evaluated by fitting the Raman spectra with Voigt peaks using the Levenberg–Marquardt algorithm. The intensity ratio of all the Raman-mode bands was evaluated by using the integrated intensity of the fitting peaks.

The sample morphology was investigated by SEM (FEI Helios 600 NanolabDualBeam), AFM (Bruker Dimension Icon microscope) operating in Tapping Mode, and ultra-high-vacuum, room-temperature STM (VT Omicron) working in constant current mode with electrochemically-etched W tips.

## 3. Results

### 3.1 XPS, Raman and SEM measurements

Figure 1 shows the XPS, Raman and SEM data of graphene films grown at 910, 920 and 930 °C for different deposition times $t_D$. The results of the quantitative analysis of the XPS and Raman spectra are reported in Table 1. As evident in the high-resolution XPS spectra reported in Fig. 1(a), all the samples show the $C_{1s}$ asymmetric line-shape peaked at 284.4 eV, typical of graphene. All the spectra are well fitted to a single component with a Doniach-Sunjic profile, with no evidence of the lower-energy spectral component associated to the carbon precursor phase of graphene observed for shorter growth time at 930 °C [27]. Moreover, since no graphene multilayer domains are evidenced by SEM measurements [Fig. 1(d-g)], the ratio $\rho$ evaluated from the XPS data and reported in the Table 1 represents the C surface coverage [27]. Interestingly, for $t_D = 60$ min, the $\rho$ value depends very slightly on the temperature, being close to 1 for all the investigated samples. This indicates that most of the Ge surface is covered by graphene. Notice that, by doubling the deposition time at $T_D$=910 °C, the surface coverage increases by less than 10%. This behaviour departs significantly from that observed on the graphene samples deposited in the same growth conditions at $T_D$= 930 °C, for which a deposition time of 120 min produces a second layer graphene covering 45 % of the surface [14].

Further insight into the properties of the graphene films can be obtained from the Raman investigation. As shown in Fig. 1(b) and well-matching the XPS data, all the spectra exhibit the main graphene features, i.e. 2D and G bands located at ~2700 and ~1600 cm⁻¹. The D (~1350 cm⁻¹) and D'(~1635 cm⁻¹) peaks, due to intervalley and intravalley resonant scattering processes induced by defects, are also clearly evident in the spectra of samples grown at 910 and 920 °C. Conversely, these features are absent or negligible for the reference sample deposited at $T_D$= 930 °C.

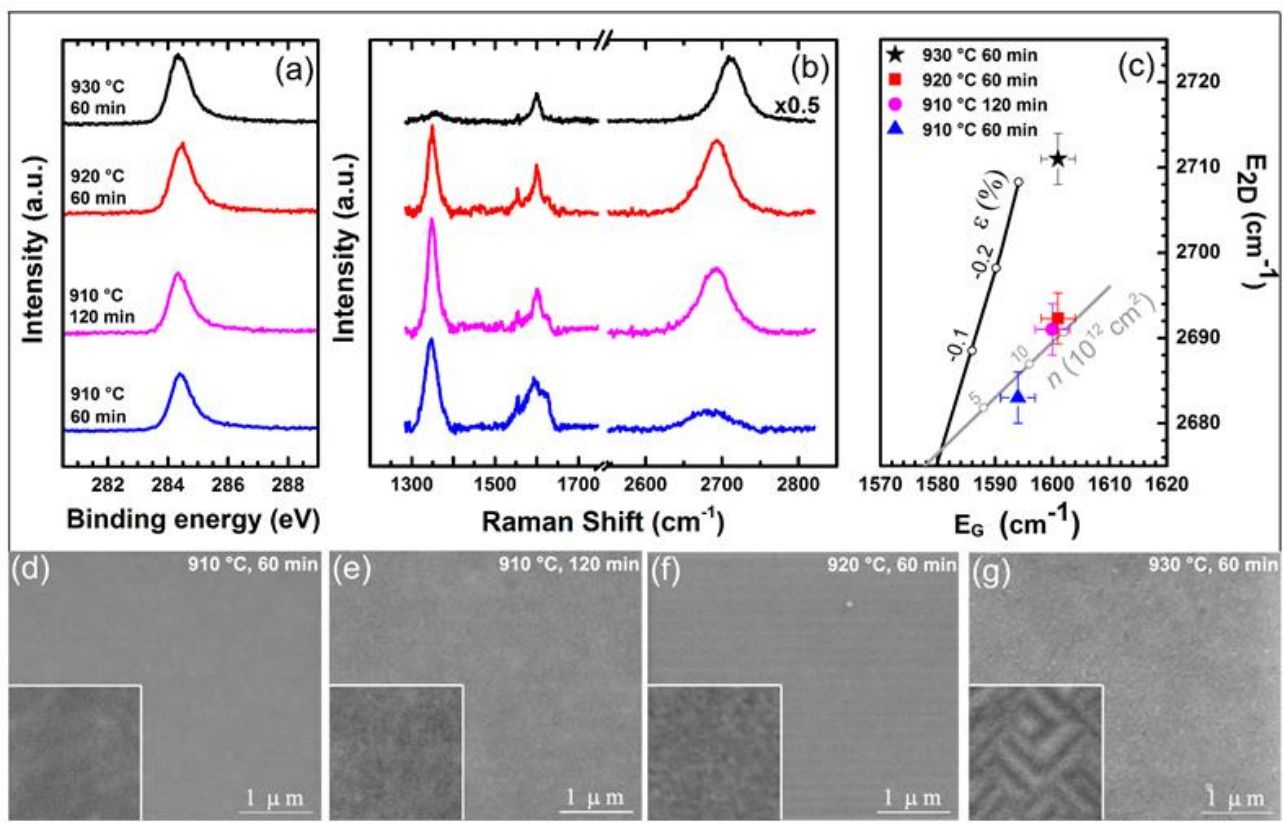

***Fig. 1.*** *(a) High resolution $C_{1s}$ core level and (b) Raman spectra of graphene films deposited at different temperature and deposition time. (c) Plot of the 2D vs G-band energies. ε is the strain and n the charge density. Straight lines indicate $E_{2D}$ vs $E_G$ relationship for strained undoped (black line, n = 0) and unstrained n-doped (grey line, ε = 0) graphene. The cross of the two lines is at the expected 2D and G positions for suspended freestanding single-layer graphene (neutrality point). The error bars represent the dispersion of the experimental values across different regions of the samples. (d-g) SEM images of the graphene samples showing no evidence of bilayer islands. In the insets 500×500 nm² images acquired at higher magnification are shown.*

| $T_D$ (°C) | $t_D$ (min) | $\rho$ (graphene monolayers) | $E_G$ (cm⁻¹) | $E_{2D}$ (cm⁻¹) | $\Gamma_{2D}$ (cm⁻¹) | $I_{2D}/I_G$ | $I_D/I_G$ | $I_{D'}/I_G$ |
|---|---|---|---|---|---|---|---|---|
| 910 | 60 | 0.81 | 1594 | 2683 | 60 | 0.4 | 1.9 | 0.5 |
| 910 | 120 | 0.86 | 1600 | 2691 | 49 | 1.5 | 2.7 | 0.2 |
| 920 | 60 | 0.88 | 1601 | 2692 | 45 | 1.8 | 2.0 | 0.2 |
| 930 | 60 | 0.97 | 1602 | 2711 | 36 | 3.6 | <0.3 | - |

***Table 1***. *Quantitative analysis of XPS and Raman spectra.*

The Raman spectra vary significantly as a function of the deposition temperature, demonstrating that the graphene quality is deeply affected by temperature, despite the relatively narrow range considered. The spectrum of the sample deposited for 60 min at the lowest temperature (910 °C) exhibits the broadest 2D peak and the lowest $I_{2D}/I_G$ intensity ratio, both demonstrating a poor crystalline quality of the graphene film [29-31]. Indeed, the high intensity of the D and D' modes indicates a high concentration of defects [32]. The doubling of the deposition time produces the raising and narrowing of the 2D band intensity. The observed increase of the $I_D/I_G$ ratio can be related to the presence of hydrogen that can induce the formation of defects in a longer growth process. Indeed the increase of the $I_D/I_G$ intensity ratio were already observed on graphene grown on copper when a large amount of $H_2$ was employed during the growth process [33]. Similarly, the increase of temperature produces a narrowing of the 2D band which reaches its minimum for the sample deposited at $T_D$= 930 °C, where $\Gamma_{2D}$ =36 cm$^{-1}$. Since no contribution to the 2D width from multilayer graphene domains is present, the larger values of $\Gamma_{2D}$ at $T_D$< 930 °C with respect to the reference single-layer graphene can only be due to strain variation on the size-scale of the laser spot [34] or a residual contribution of defects to the 2D band [29]. Note that the $I_D/I_G$ intensity ratio strongly decreases only at 930°C.

The strain and doping levels, averaged on the size of the laser spot, can be inferred from the analysis of the 2D vs G band energies [35]. It is noteworthy that the average strain in the graphene films grown at $T_D$<930 °C is negligible [Fig. 1(c)]. By contrast, the graphene layer deposited at $T_D$= 930 °C shows a compressive strain of about 0.3%, in line with the literature [14, 15, 17]. As typically found for graphene/Ge(001) [14, 17, 22, 24], all the samples show a non-vanishing average charge density $n$ of the order of 10$^{13}$ cm$^{-2}$.

*3.2 Morphological investigation*

The XPS, Raman and SEM investigations show that an almost-complete graphene monolayer is formed between 910 and 930 °C and for $t_D$= 60 min. This implies that the efficiency of catalysis and graphene formation is not altered within this narrow temperature range. Surprisingly, however, the Raman analysis also shows that such a small variation in the deposition temperature dramatically affects the quality of the graphene adlayer, which degrades significantly at a temperature less than 30 degrees below the Ge melting point (nominally 937.4 °C at atmospheric pressure [36]). In order to gain more insights into the mechanism leading to this abrupt change with temperature, we studied the morphological and structural evolution of the sample surface at different size scales. Nearing the Ge melting point, a variation in the growth temperature of few tens of degrees determines significant transformations in the mesoscale morphology, these being observable in the AFM images reported in Fig. 2(a-c). For the same growth time $t_D$= 60 min, the surface lacks any evident mesoscale ordering for $T_D$= 910 °C [Fig 2(a)], while it breaks up into a regular faceting pattern at $T_D$= 930 °C which is the distinctive fingerprint of high-quality graphene on Ge(001) [Fig. 2(c)]. Note that the faceting develops only when the Ge substrate is heated at 930 °C in presence of methane. In fact, an unaltered staircase of steps and terraces along the (001) plane is observed for a test growth performed at the same temperature but in $H_2$/Ar atmosphere only without $CH_4$ [inset in Fig. 2(c)]. The faceting consists of two families of {10$L$} facets aligned along <010> directions and with a relative azimuthal orientation of 90° which defines a texture of Ge hut clusters with average height of 4 nm extending over the whole substrate surface. In line with previous observations [14, 17, 22-25], the misorientation angle of the hut-cluster facets, with respect to the (001) plane, ranges between 5° and 8° (corresponding to $L$= 7÷10) and is therefore shallower than that typically observed for strain-induced faceting of the Ge(001) films [37, 38] and Ge quantum dots [39-44]. Confirming the reproducibility and control over temperature, at the intermediate temperature of 920 °C [Fig. 2(b)], we observe a faint, initial structuring of the substrate which is evident in the appearance of shallow elongated mounds with an average height of about 1 nm and aligned preferentially along the <010> directions, i.e. along the image diagonals. The morphological changes induced by the temperature are accompanied by a progressive increase in the root-mean square roughness between 910 and 930 °C [black markers in Fig. 2(e)]. We remark that this roughening is strictly related to the growth of a graphene film, since in the test-growth Ge sample annealed up to 930 °C [inset Fig. 2(c)] the RMS roughness is equal to 1.4 Å [cyan marker in Fig. 2(e)] and is determined by the height of Ge(001) steps. Visually, the increased height modulation of the Ge substrate covered by the graphene layer with temperature is clear from cross-sectional line profiles taken along the horizontal axis of the AFM images [Fig. 2(f)]. Over a profile length of 500 nm, the maximum peak (or valley) from the average profile line is less than 0.5 nm at 910 °C, while it is increased to about 1 nm at 920 °C and grows even further at 930 °C, approaching almost 2 nm. We observe that, at a constant temperature of $T_D$= 910 °C, doubling the growth time also results in a slight increase of the roughness, even though the overall effect on the substrate morphology is weaker [Fig. 2(e-f)].

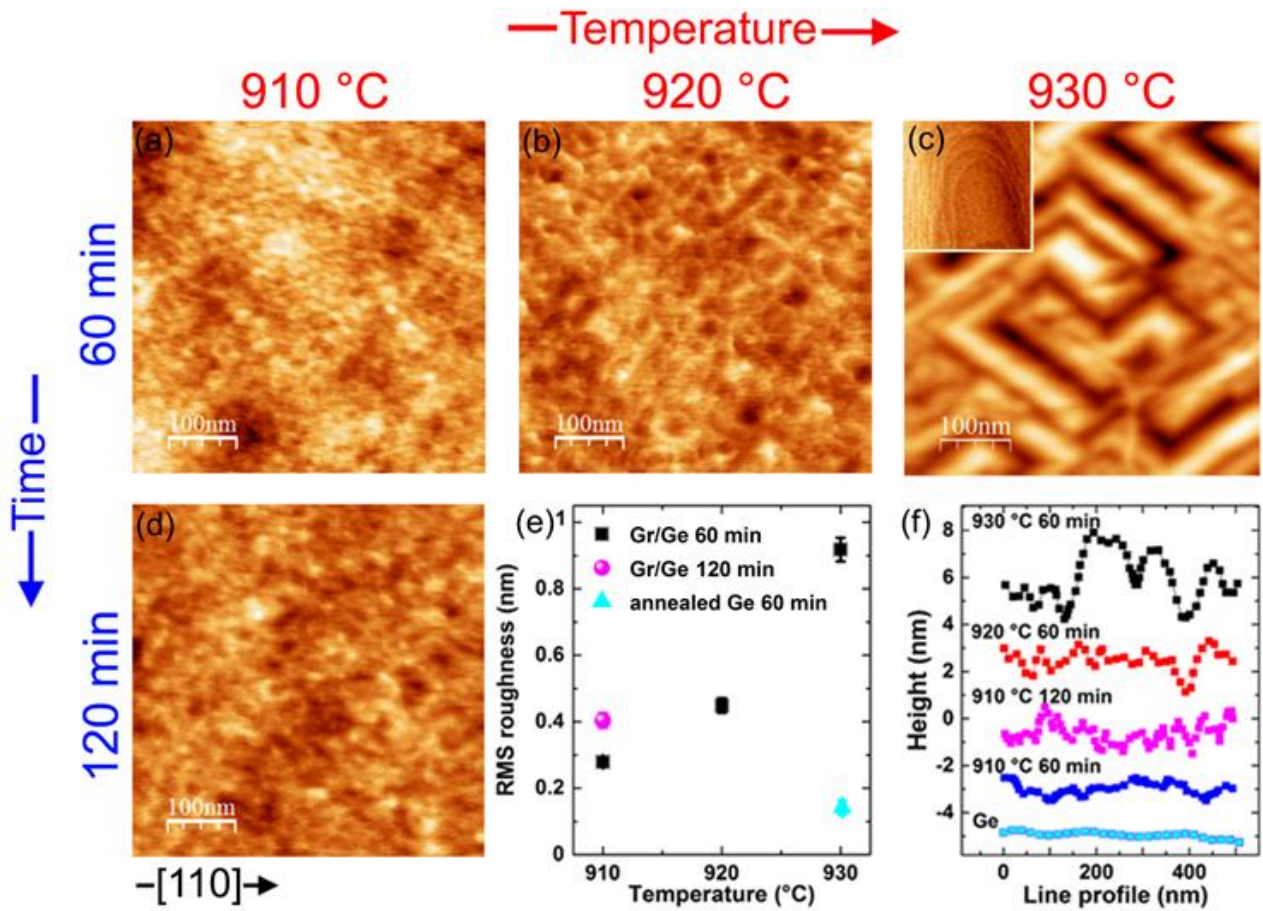

**_Fig. 2_**_. AFM images and corresponding topographic parameters of graphene/Ge(001) samples as a function of temperature and deposition time. (a) $T_D$= 910 °C, $t_D$= 60 min. (b) $T_D$= 920 °C, $t_D$= 60 min. (c) $T_D$= 930 °C, $t_D$= 60 min. (d) $T_D$= 910 °C, $t_D$= 120 min. The arrow indicates the [110] direction. In the inset of panel (c), we report the AFM topography of a test-growth Ge sample annealed to 930 °C for 60 min in $H_2$/Ar without methane. (e) Root-mean-square (RMS) roughness obtained from the_

*AFM images of the different graphene samples. As reference, the RMS roughness measured on the annealed Ge sample is also reported. (f) Height profile taken on the horizontal axis of the AFM images of the graphene samples and on the annealed Ge sample.*

A close-up on the structural changes induced by the temperature and growth time is obtained by STM measurements. Each column in Fig. 3 shows data for the same sample at different length-scales: The first two rows represent a length scale at which the STM is most sensitive to the topography of the Ge substrate, while the two remaining rows are focused on the morphology of the graphene layer emerging at the atomic scale. The horizontal stacking of the images within the same row shows the effect of temperature (second towards fourth column) or growth time (second towards first column) at a given length scale. We first focus on the structural changes occurring on the Ge substrate, i.e. on the first two rows. The increased resolution of STM with respect to AFM makes it possible to highlight the changes occurring at the mesoscale. The evolutionary path at increasing temperature indicates that the build-up of faceting is preceded by a re-organization of the texture of the substrate in which an untextured and homogeneously corrugated surface observed at $T_D$= 910 °C evolves first into a rugged surface of alternating shallow mounds and valleys ($T_D$= 920 °C); then, at $T_D$= 930 °C, the mounds grow in height (from 1.4 ±0.2 nm at 920 °C to 4.5 ±0.2 nm, in close agreement with the AFM results) and develop well-defined facets, thus evolving into the hut-cluster network which patterns the whole surface. STM data also confirm that an increase in the growth time drives a markedly diminished structuring and ordering effect on the substrate than that observed for the increase in the growth temperature. Namely, for $t_D$= 120 min at $T_D$= 910 °C, we observe the formation of plateaus and platelets which replace the round-shape grains detected for $t_D$= for 60 min (see the second and first columns in Fig. 3). As visually clear in Fig. 3, these structures are markedly smaller than the mounds and huts formed at $T_D$= 920 °C and 930 °C, respectively.

By merging morphological information from different length-scales, STM imaging reveals a subtle interplay between the mesoscale topography of the Ge substrate and the atomic-scale structure of the graphene adlayer. At a size scale of a few nanometers (the last two rows in Fig. 3), high-resolution STM images detect the arrangement of carbon atoms in the graphene sheet. A visual impression of the different structural corrugations of the graphene layer, as a function of temperature or growth time, is provided by the sequential set of three-dimensional STM images displayed on the third row of Fig. 3. On the well-defined Ge facets formed at $T_D$= 930 °C, the graphene layer is flat (with a peak-to valley depth < 2 Å) and shows an almost perfect crystalline order of the honeycomb lattice (See last column in Fig. 3). The high quality of the graphene film grown at 930 °C is further confirmed by the sharp hexagonal pattern observed in the Fast-Fourier Transform (FFT) of the corresponding atomic-resolution STM image (See inset). Conversely, when the faceting vanishes at lower temperatures, the irregular Ge surface provides an uneven template for the graphene sheet, resulting in the formation of multiple ripples (of about 4 Å in height) and, for $T_D$= 910 °C and $t_D$= 60 min, also in the formation of larger-amplitude wrinkles with a maximum height >1 nm. At the atomic scale, lattice deformations and point defects such as C vacancies are visible in high-resolution STM images (see the first three panels of the last row of Fig. 3). Furthermore, matching the results obtained on different graphene systems [45-48], we observe that the increased local curvature, revealed at $T_D$= 910 °C and $t_D$= 60 min, produces local perturbations strong enough to break the six-fold symmetry of the carbon lattice in graphene. On this sample, as evident in Fig. 4(a), regions with the regular honeycomb lattice (central part of the top inset) are alternated with areas where the honeycomb pattern is strongly distorted (bottom inset). In some locations of Fig. 4(a), the honeycomb cannot be seen, whereas we observe atomic rows with a separation compatible with the second nearest-neighbour distance in graphene, i.e. 0.246 nm. This value corresponds to the periodicity of a single triangular sublattice of graphene, indicating that only one sublattice can be observed. This is consistent with characteristic *three-for-six* defects, typically associated with corrugated graphene films [45, 47, 48]. The origin of such motifs is attributed to a local distortion of the $sp^2$ bonding configuration induced by the curvature which lifts three of the C atoms up from the hexagon plane. Indeed, the presence of a strong curvature in the graphene layer is indicated by the meandering of the atomic rows, observed in Fig. 4(a). In Figs. 4(b) and 4(c), we compare the FFT maps acquired on different areas of this sample, showing, respectively, the smallest (b) and the largest corrugation (c). On the former, the FFT map shows the hexagonal pattern of the honeycomb lattice [Fig. 4(b)], although the broader spots with respect to $T_D$= 930 °C (inset Fig. 3) confirm that the graphene film is significantly more corrugated than at higher growth temperature. Where the graphene film is locally strongly wrinkled (see Fig.3, last two rows, second column), the six-fold spots of the FFT have low intensity and additional features appear close to the center of the map at a wavelength matching the wrinkle pitch [Fig. 4(c)].

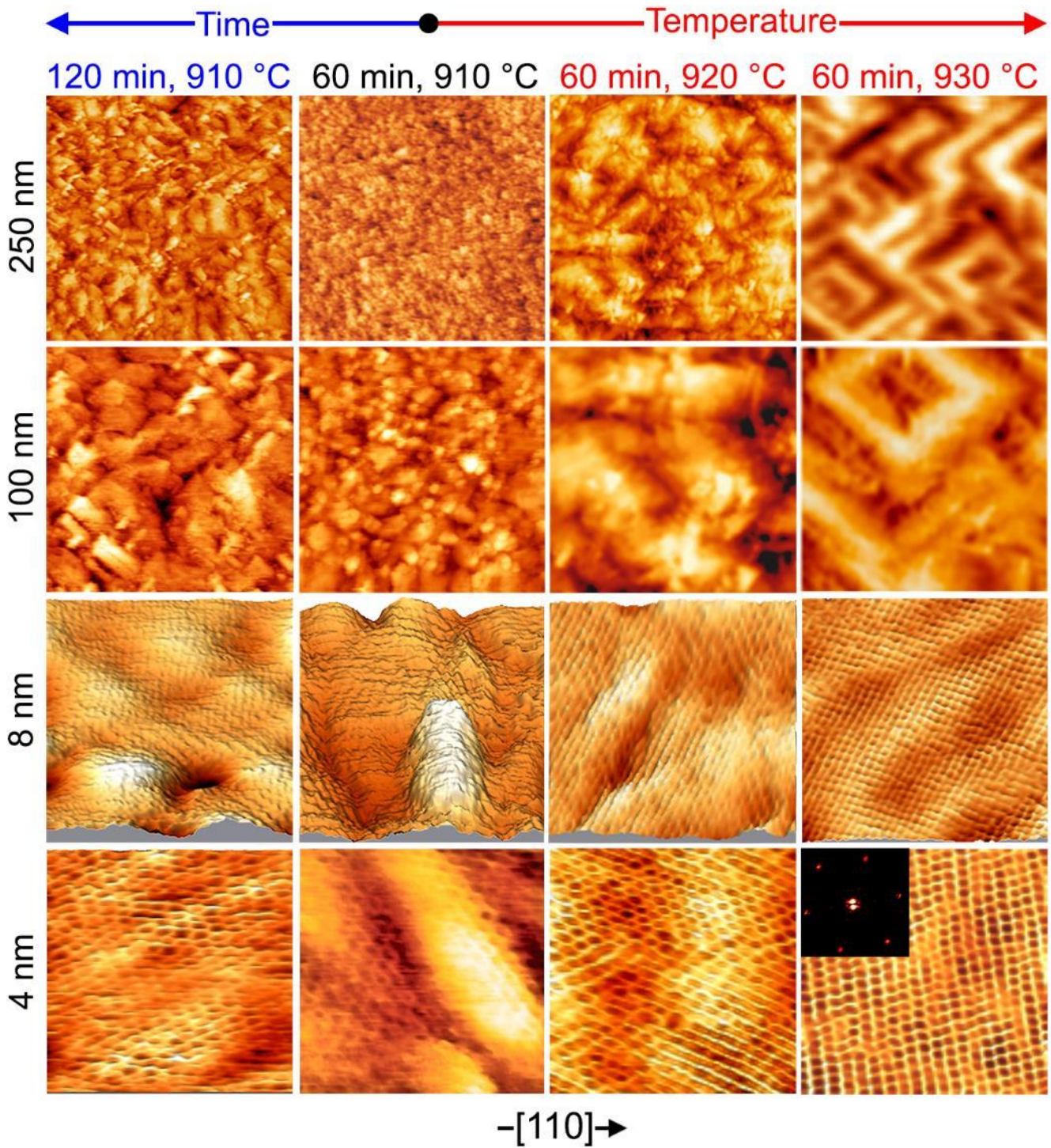

***Fig. 3.*** *STM images of graphene/Ge(001) samples as a function of temperature and deposition time. Each column corresponds to a different sample grown with the parameters reported above. The temperature increases going from the second to the fourth column, while the deposition time increases going from the second to the first column. The image size changes moving vertically along different rows of a given column. The arrow indicates the [110] direction. The third row shows a comparison of 3D topographies displayed with the same z-axis range 0-12 Å. The morphological evolution shown is repeatedly observed on different areas of the samples. The inset shows the FFT of*

*the high-quality honeycomb lattice observed for $T_D$=930 °C and $t_D$=60 min. STM images are taken with the following tunneling parameters U= 50 mV, I=1.8 nA.*

The different corrugations of the graphene layer, induced by the temperature-dependent topography of the Ge template, severely affect the electronic properties of graphene, as shown by spatially-resolved scanning tunneling spectroscopy (STS) [Fig. 4(d, e)]. As already reported [17, 22, 26], the flat graphene monolayer grown on the {10L} facets at $T_D$= 930 °C shows a quasi-free-standing behaviour demonstrated by a linear differential conductance [Fig. 4(d)]. This suggests that on the flat Ge surface of the facets the Van-der Walls interaction does not impact on the local density of states. Only locally, on the valley locations of the faceting pattern (i.e. in between the facets of two adjacent hut-clusters), a quadratic electron local density of states (LDOS) was reported, suggesting an higher graphene-Ge interaction [17, 26]. Conversely, we find that a parabolic tunneling conductance is measured everywhere at lower growth temperature ($T_D$= 910 °C) [Fig. 4(e)], where the Ge substrate is not faceted and a high density of ripples and wrinkles is present in the graphene adlayer. This finding matches results obtained for hydrogen-intercalated [49] or annealed [26] graphene/Ge in which the degradation of the Ge nanofaceting is accompanied by distortions in the graphene structure and a significant deviation of the tunneling conductance from the free-standing behaviour, with a lower overall conductance and a non-linear $dI/dV$ ratio.

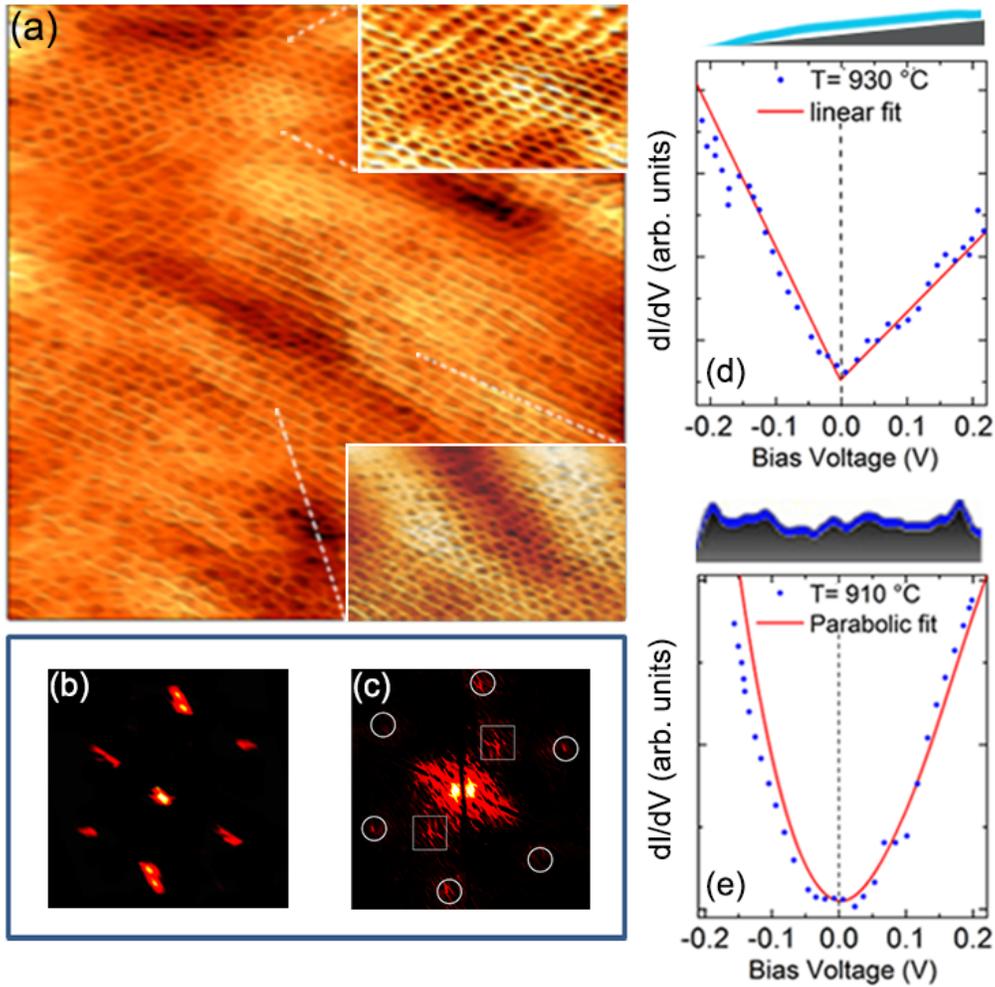

**_Fig. 4._** **(a) High-resolution STM images (13x 13) nm² of a graphene/Ge(001) sample growth at $T_D=$ 910 °C for $t_D=$ 60 min. The blow-ups show enlarged views of areas with different corrugation. (b, c) FFTs obtained in different areas of the sample: (b) on a region without wrinkles, (c) on a wrinkled area where shallow six-fold spots (circles) coexist with long-wavelength features (squares) corresponding to the wrinkle pitch. (d, e) STS spectroscopy curves obtained (d) on the graphene sample grown $T_D=$930 °C and for $t_D=$ 60 min and (e) on the graphene sample grown at $T_D=$ 910 °C for $t_D=$ 60 min. The curves shown are averaged over 100 acquisitions and on an area of 20x 20 nm².The upper inserts show a pictorial view of conforming graphene films at the different growth temperature.**

## 4. Discussion

Our analysis shows that two correlative effects coexist: On the one hand, the Ge surface, under the graphene layer, corrugates at the nanometer-scale below 930 °C and, in turn, the graphene bulks up to form nano-wrinkles and ripples and develops a larger density of point defects. On the other hand, at $T_D=$ 930 °C the absence of short-wavelength corrugations (at the scale of C-C bonds) on the Ge facets results in a high-quality graphene layer flat on the scale of the facet size. In addition, any modelling framework aiming to clarify the interplay between graphene and Ge needs to incorporate three key observations: *i)* the dramatic changes in the Ge topography associated to the extended faceting at $T_D$= 930 °C occur only when Ge is heated up in the presence of a C-precursor gas; *ii)* these morphological alterations take place within an extremely narrow temperature range; *iii)* the efficiency of catalysis and graphene formation is not altered within this narrow temperature range.

Showing similar catalytic activity and solubility, the well-known graphene/Cu system provides an instructive comparison to gain insights into the process acting on Ge [14]. On Cu, temperature-dependent studies show that striking alterations of the substrate topography and of the graphene morphology appear at the onset of Cu surface melting, as a consequence of the increased dynamics in the melted layer [50-52]. Surface melting consists of a *quasi*-liquid outermost film which wets the underlying solid substrate and shows structural properties and mobility halfway between a solid and a liquid. On Cu, it has been shown that a quasi-liquid layer is formed at about 800 °C, i.e. more than 250 °C below the melting point. The onset of surface melting is the point where Ge diverges substantially from Cu. Like most semiconductors, Ge has a negative Hamaker constant for long-range dispersion forces, which hinders complete surface melting [53]. This means that a quasi-liquid adlayer is stabilized only in close (i.e. a few degrees) proximity to the Ge melting point [54]. We believe that the incomplete-melting behaviour of Ge(001) is at the base of the abrupt temperature dependence observed in the graphene/Ge system. The surface melted layer is formed only at $T_D$= 930 °C and its presence affects strongly the graphene growth. In particular, the weakening of the in-plane bonding in the melted layer determines a higher diffusivity and sublimation rate for the surface atoms of the substrate, with a strong impact on conformal graphene films [50, 51]. From the parallel with the Cu case, in which the growing graphene sheets provide a barrier against sublimation [51], and as also observed for graphene nanoribbons on Ge [55], it is reasonable to assume that the evaporation from the quasi-liquid Ge layer is hampered considerably under the growing graphene domains, since the Ge atoms must first diffuse outward from the core of the graphene islands onto the exposed Ge surface, in order to sublimate. We believe that this local asymmetry in the sublimation rate induced by graphene kick-starts the mesoscale transformation of the Ge surface, promoting the formation of mounds selectively on the graphene-covered areas. At $T_D$= 930°C, where sublimation is particularly severe, these height-modulation instabilities in the surface melted layer, induced by graphene, reach the critical aspect ratio for faceting along vicinal orientations of the (001) face, corresponding to local surface-energy minima of Ge [56]. Since the sublimation rate and atom diffusivity drops abruptly when surface melting of Ge is quenched, we observe, accordingly, that Ge faceting vanishes at 910 °C. On the other hand the Ge corrugated morphology observed in samples grown at 910 °C in which a defected graphene sheet cover almost completely the substrate, suggests that the catalysis process yielding to the formation of graphene increases the structural disorder of the substrate surface at the nanoscale [57]. On metal templates, short-wavelength corrugations in graphene arise from the tendency of the growing graphene layer to maintain locally the optimum distance from the substrate for van der Waals interactions [58]. Consequently, the graphene sheets bend to follow the optimal contour of interaction potential. Clearly, this correlative effect between corrugations at atomic scale in graphene and in the substrate underneath also acts on Ge. Indeed, in order to adapt to this uneven substrate profile [diagram in Fig. 4(e)], the graphene sheet grown at $T$< 930 °C develops a pronounced curvature with a high concentration of ripples and nano-wrinkles, as observed in the STM experiments. Distortions in the graphene structure indicate the presence of strong interactions between graphene and Ge [26] which, electronically, result in the parabolic tunneling conductance probed by STS. Conversely, in the presence of surface melting at 930 °C, the quasi liquid Ge layer, being highly-mobile, smoothens out the local corrugations of the surface, so favoring its faceting and thus preventing the curvature in graphene. At the scale of the facet size (typically, several tens of nanometers), the Ge substrate provides an atomically-flat and electronically homogeneous support that allows the graphene layer to assemble and self-organize with a reduced interference from the support [diagram in Fig. 4(d)] as confirmed by the linear behavior of the tunneling conductance. The visually-clear improvement of crystalline graphene quality revealed by high-resolution STM and the abrupt decreases of D and D' peaks at increasing temperature in the narrow range of temperature below the melting both demonstrate that the quasi liquid substrate favors the healing of defects in graphene, as already suggested in Refs. [57, 59] for the growth of graphene on metals.

The high-density of defects and strain variations at the atomic-scale, induced by the short-wavelength surface corrugation at T< 930 °C detected by STM, are responsible for the widening of the 2D Raman band reported in Table 1. Since these local strains average out at the scale of the laser spot, a vanishing average strain is obtained from the $E_{2D}$ $vs$ $E_G$ analysis. Conversely, on the graphene grown on the atomically-flat facets at $T$= 930 °C, the Raman investigation suggests the presence of a compressive strain of 0.3% which is in line with previous published values attributed to the different thermal expansion coefficient of graphene and Ge or to the development of the nanofaceting [17, 22]. The negligible strain measured in samples grown at 910 and 920 °C suggests a minor role played by the former mechanism.

Finally, we suggest that the enhanced mobility of Ge atoms in the surface-melting regime can explain the nucleation of a second graphene layer in a layer-by-layer regime at $T_D$= 930 °C [14], in contrast to the self-limited graphene growth observed here at lower temperature when the Ge surface is solid. Indeed, in the quasi-liquid layer, Ge atoms may easily diffuse upwards to the surface of the first graphene layer by finding their way through vacancy and edge boundaries and promoting the catalysis of a second graphene layer.

## 5. Conclusions

By a fine tuning of the growth temperature in close proximity of the Ge melting point, we highlight a clear discontinuity in the graphene/Ge(001) system at $T_D$= 930 °C, in terms of growth mode, defect density and structural/electronic properties of the graphene layer. We identify this discontinuity as the onset of surface melting in the Ge(001) substrate which develops a quasi-liquid outermost layer only within a few degrees from the melting point. Our findings demonstrate that the presence of this quasi-liquid layer is fundamental to improve the quality of graphene and that an accurate control over the growth temperature is mandatory to achieve, in a reproducible way, a high-quality graphene monolayer on Ge(001).

**Conflicts of interest**

There are no conflicts to declare.

**Acknowledgements**

The Lime Laboratory of Roma Tre University is acknowledged for technical support. Funding from the European Union's Horizon 2020 research and innovation program under grant agreement No. 696656 - GrapheneCore1 is acknowledged.

**References**

[1] F. Schwierz, H. Wong, J.J. Liou, Nanometer CMOS, Pan Stanford, Singapore, 2010.

[2] International Roadmap for Devices and Systems, www.irds.ieee.org2017.

[3] A. Pospischil, M. Humer, M.M. Furchi, D. Bachmann, R. Guider, T. Fromherz, T. Mueller, CMOS-compatible graphene photodetector covering all optical communication bands, Nat. Photonics 7 (2013) 892.

[4] F. Withers, O. Del Pozo-Zamudio, A. Mishchenko, A.P. Rooney, A. Gholinia, K. Watanabe, T. Taniguchi, S.J. Haigh, A.K. Geim, A.I. Tartakovskii, K.S. Novoselov, Light-emitting diodes by band-structure engineering in van der Waals heterostructures, Nat. Mater. 14 (2015) 301.

[5] J. Gosciniak, D.T.H. Tan, Theoretical investigation of graphene-based photonic modulators, Scientific Reports 3 (2013) 1897.

[6] S. Goossens, G. Navickaite, C. Monasterio, S. Gupta, J.J. Piqueras, R. Pérez, G. Burwell, I. Nikitskiy, T. Lasanta, T. Galán, E. Puma, A. Centeno, A. Pesquera, A. Zurutuza, G. Konstantatos, F. Koppens, Broadband image sensor array based on graphene–CMOS integration, Nat. Photonics 11 (2017) 366.

[7] M. Mittendorff, S. Winnerl, J. Kamann, J. Eroms, D. Weiss, H. Schneider, M. Helm, Ultrafast graphene-based broadband THz detector, Appl. Phys. Lett. 103(2) (2013) 021113.

[8] C.-H. Liu, Y.-C. Chang, T.B. Norris, Z. Zhong, Graphene photodetectors with ultra-broadband and high responsivity at room temperature, Nat. Nanotechnol. 9 (2014) 273.

[9] V. Sorianello, M. Midrio, G. Contestabile, I. Asselberghs, J. Van Campenhout, C. Huyghebaert, I. Goykhman, A.K. Ott, A.C. Ferrari, M. Romagnoli, Graphene–silicon phase modulators with gigahertz bandwidth, Nat. Photonics 12(1) (2018) 40-44.

[10] K.S. Novoselov, V.I. Fal'ko, L. Colombo, P.R. Gellert, M.G. Schwab, K. Kim, A roadmap for graphene, Nature 490 (2012) 192.

[11] G. Wang, M. Zhang, Y. Zhu, G. Ding, D. Jiang, Q. Guo, S. Liu, X. Xie, P.K. Chu, Z. Di, X. Wang, Direct Growth of Graphene Film on Germanium Substrate, Scientific Reports 3 (2013) 2465.

[12] J.-H. Lee, E.K. Lee, W.-J. Joo, Y. Jang, B.-S. Kim, J.Y. Lim, S.-H. Choi, S.J. Ahn, J.R. Ahn, M.-H. Park, C.-W. Yang, B.L. Choi, S.-W. Hwang, D. Whang, Wafer-Scale Growth of Single-Crystal Monolayer Graphene on Reusable Hydrogen-Terminated Germanium, Science 344(6181) (2014) 286-289.

[13] I. Pasternak, M. Wesolowski, I. Jozwik, M. Lukosius, G. Lupina, P. Dabrowski, J.M. Baranowski, W. Strupinski, Graphene growth on Ge(100)/Si(100) substrates by CVD method, Scientific Reports 6 (2016) 21773.

[14] A.M. Scaparro, V. Miseikis, C. Coletti, A. Notargiacomo, M. Pea, M. De Seta, L. Di Gaspare, Investigating the CVD Synthesis of Graphene on Ge(100): toward Layer-by-Layer Growth, ACS Applied Materials & Interfaces 8(48) (2016) 33083-33090.

[15] J. Dabrowski, G. Lippert, J. Avila, J. Baringhaus, I. Colambo, Y.S. Dedkov, F. Herziger, G. Lupina, J. Maultzsch, T. Schaffus, T. Schroeder, M. Kot, C. Tegenkamp, D. Vignaud, M.C. Asensio, Understanding the growth mechanism of graphene on Ge/Si(001) surfaces, Scientific Reports 6 (2016) 31639.

[16] J. Tesch, E. Voloshina, M. Fonin, Y. Dedkov, Growth and electronic structure of graphene on semiconducting Ge(110), Carbon 122 (2017) 428-433.

[17] I. Pasternak, P. Dabrowski, P. Ciepielewski, V. Kolkovsky, Z. Klusek, J.M. Baranowski, W. Strupinski, Large-area high-quality graphene on Ge(001)/Si(001) substrates, Nanoscale 8(21) (2016) 11241-11247.

[18] C.D. Mendoza, P.G. Caldas, F.L. Freire, M.E.H. Maia da Costa, Growth of single-layer graphene on Ge (1 0 0) by chemical vapor deposition, Appl. Surf. Sci. 447 (2018) 816-821.

[19] K. Kim, J.-Y. Choi, T. Kim, S.-H. Cho, H.-J. Chung, A role for graphene in silicon-based semiconductor devices, Nature 479 (2011) 338.

[20] T. Hallam, N.C. Berner, C. Yim, G.S. Duesberg, Strain, Bubbles, Dirt, and Folds: A Study of Graphene Polymer-Assisted Transfer, Advanced Materials Interfaces 1(6) (2014) 1400115.

[21] G. Lupina, J. Kitzmann, I. Costina, M. Lukosius, C. Wenger, A. Wolff, S. Vaziri, M. Östling, I. Pasternak, A. Krajewska, W. Strupinski, S. Kataria, A. Gahoi, M.C. Lemme, G. Ruhl, G. Zoth, O. Luxenhofer, W. Mehr, Residual Metallic Contamination of Transferred Chemical Vapor Deposited Graphene, ACS Nano 9(5) (2015) 4776-4785.

[22] B. Kiraly, R.M. Jacobberger, A.J. Mannix, G.P. Campbell, M.J. Bedzyk, M.S. Arnold, M.C. Hersam, N.P. Guisinger, Electronic and Mechanical Properties of Graphene–Germanium Interfaces Grown by Chemical Vapor Deposition, Nano Lett. 15(11) (2015) 7414-7420.

[23] K.M. McElhinny, R.M. Jacobberger, A.J. Zaug, M.S. Arnold, P.G. Evans, Graphene-induced Ge (001) surface faceting, Surf. Sci. 647 (2016) 90-95.

[24] M. Lukosius, J. Dabrowski, J. Kitzmann, O. Fursenko, F. Akhtar, M. Lisker, G. Lippert, S. Schulze, Y. Yamamoto, M.A. Schubert, H.M. Krause, A. Wolff, A. Mai, T. Schroeder, G. Lupina, Metal-Free CVD Graphene Synthesis on 200 mm Ge/Si(001) Substrates, ACS Applied Materials & Interfaces 8(49) (2016) 33786-33793.

[25] R.M. Jacobberger, B. Kiraly, M. Fortin-Deschenes, P.L. Levesque, K.M. McElhinny, G.J. Brady, R. Rojas Delgado, S. Singha Roy, A. Mannix, M.G. Lagally, P.G. Evans, P. Desjardins, R. Martel, M.C.

Hersam, N.P. Guisinger, M.S. Arnold, Direct oriented growth of armchair graphene nanoribbons on germanium, Nature Communications 6 (2015) 8006.

[26] P. Dabrowski, M. Rogala, I. Pasternak, J. Baranowski, W. Strupinski, M. Kopciuszynski, R. Zdyb, M. Jalochowski, I. Lutsyk, Z. Klusek, The study of the interactions between graphene and Ge(001)/Si(001), Nano Research 10(11) (2017) 3648-3661.

[27] L. Di Gaspare, A.M. Scaparro, M. Fanfoni, L. Fazi, A. Sgarlata, A. Notargiacomo, V. Miseikis, C. Coletti, M. De Seta, Early stage of CVD graphene synthesis on Ge(001) substrate, Carbon 134 (2018) 183-188.

[28] G. Lippert, J. Dąbrowski, T. Schroeder, M.A. Schubert, Y. Yamamoto, F. Herziger, J. Maultzsch, J. Baringhaus, C. Tegenkamp, M.C. Asensio, J. Avila, G. Lupina, Graphene grown on Ge(001) from atomic source, Carbon 75 (2014) 104-112.

[29] E.H. Martins Ferreira, M.V.O. Moutinho, F. Stavale, M.M. Lucchese, R.B. Capaz, C.A. Achete, A. Jorio, Evolution of the Raman spectra from single-, few-, and many-layer graphene with increasing disorder, Phys. Rev. B 82(12) (2010) 125429.

[30] P. Venezuela, M. Lazzeri, F. Mauri, Theory of double-resonant Raman spectra in graphene: Intensity and line shape of defect-induced and two-phonon bands, Phys. Rev. B 84(3) (2011) 035433.

[31] L.G. Cançado, A. Jorio, E.H.M. Ferreira, F. Stavale, C.A. Achete, R.B. Capaz, M.V.O. Moutinho, A. Lombardo, T.S. Kulmala, A.C. Ferrari, Quantifying Defects in Graphene via Raman Spectroscopy at Different Excitation Energies, Nano Lett. 11(8) (2011) 3190-3196.

[32] A. Ferrari, D. Basko, Raman spectroscopy as a versatile tool for studying the properties of graphene, Nat Nano 8(4) (2013) 235-246.

[33] M. Losurdo, M.M. Giangregorio, P. Capezzuto, G. Bruno, Graphene CVD growth on copper and nickel: role of hydrogen in kinetics and structure, Phys. Chem. Chem. Phys. 13(46) (2011) 20836-20843.

[34] C. Neumann, S. Reichardt, P. Venezuela, M. Drögeler, L. Banszerus, M. Schmitz, K. Watanabe, T. Taniguchi, F. Mauri, B. Beschoten, S.V. Rotkin, C. Stampfer, Raman spectroscopy as probe of nanometre-scale strain variations in graphene, Nature Communications 6 (2015) 8429.

[35] J.E. Lee, G. Ahn, J. Shim, Y.S. Lee, S. Ryu, Optical separation of mechanical strain from charge doping in graphene, Nature Communications 3 (2012) 1024.

[36] S.M. Sze, Physics of Semiconductor Devices, Wiley and Sons, New York, 1981.

[37] L. Fazi, C. Hogan, L. Persichetti, C. Goletti, M. Palummo, A. Sgarlata, A. Balzarotti, Intermixing and buried interfacial structure in strained Ge/Si(105) facets, Phys. Rev. B 88(19) (2013) 195312.

[38] L. Persichetti, A. Sgarlata, G. Mattoni, M. Fanfoni, A. Balzarotti, Orientational phase diagram of the epitaxially strained Si(001): Evidence of a singular (105) face, Phys. Rev. B 85(19) (2012) 195314.

[39] G. Capellini, M. De Seta, F. Evangelisti, Influence of the growth parameters on self-assembled Ge islands on Si(100), Mater. Sci. Eng: B 89(1) (2002) 184-187.

[40] M. De Seta, G. Capellini, L. Di Gaspare, F. Evangelisti, F. D'Acapito, Freezing shape and composition of Ge/Si(001) self-assembled islands during silicon capping, J. Appl. Phys. 100(9) (2006) 093516.

[41] F. Montalenti, A. Marzegalli, G. Capellini, D. Seta, L. Miglio, Vertical and lateral ordering of Ge islands grown on Si(001): theory and experiments, J. Phys.: Condens. Matter 19(22) (2007) 225001.

[42] L. Persichetti, A. Sgarlata, M. Fanfoni, A. Balzarotti, Heteroepitaxy of Ge on singular and vicinal Si surfaces: elastic field symmetry and nanostructure growth, J. Phys.: Condens. Matter 27(25) (2015) 253001.

[43] L. Persichetti, A. Sgarlata, M. Fanfoni, A. Balzarotti, Pair interaction between Ge islands on vicinal Si(001) surfaces, Phys. Rev. B 81(11) (2010) 113409.

[44] L. Persichetti, A. Sgarlata, M. Fanfoni, A. Balzarotti, Ripple-to-dome transition: The growth evolution of Ge on vicinal Si(1 1 10) surface, Phys. Rev. B 82(12) (2010) 121309.

[45] K. Xu, P. Cao, J.R. Heath, Scanning Tunneling Microscopy Characterization of the Electrical Properties of Wrinkles in Exfoliated Graphene Monolayers, Nano Lett. 9(12) (2009) 4446-4451.

[46] M. Ishigami, J.H. Chen, W.G. Cullen, M.S. Fuhrer, E.D. Williams, Atomic Structure of Graphene on $SiO_2$, Nano Lett. 7(6) (2007) 1643-1648.

[47] Y. Zhang, T. Gao, Y. Gao, S. Xie, Q. Ji, K. Yan, H. Peng, Z. Liu, Defect-like Structures of Graphene on Copper Foils for Strain Relief Investigated by High-Resolution Scanning Tunneling Microscopy, ACS Nano 5(5) (2011) 4014-4022.

[48] M. Lanza, Y. Wang, A. Bayerl, T. Gao, M. Porti, M. Nafria, H. Liang, G. Jing, Z. Liu, Y. Zhang, Y. Tong, H. Duan, Tuning graphene morphology by substrate towards wrinkle-free devices: Experiment and simulation, J. Appl. Phys. 113(10) (2013) 104301.

[49] J. Grzonka, I. Pasternak, P.P. Michałowski, V. Kolkovsky, W. Strupinski, Influence of hydrogen intercalation on graphene/Ge(0 0 1)/Si(0 0 1) interface, Appl. Surf. Sci. 447 (2018) 582-586.

[50] J.M. Wofford, S. Nie, K.F. McCarty, N.C. Bartelt, O.D. Dubon, Graphene Islands on Cu Foils: The Interplay between Shape, Orientation, and Defects, Nano Lett. 10(12) (2010) 4890-4896.

[51] Z.-J. Wang, G. Weinberg, Q. Zhang, T. Lunkenbein, A. Klein-Hoffmann, M. Kurnatowska, M. Plodinec, Q. Li, L. Chi, R. Schloegl, M.-G. Willinger, Direct Observation of Graphene Growth and Associated Copper Substrate Dynamics by in Situ Scanning Electron Microscopy, ACS Nano 9(2) (2015) 1506-1519.

[52] B. Deng, J. Wu, S. Zhang, Y. Qi, L. Zheng, H. Yang, J. Tang, L. Tong, J. Zhang, Z. Liu, H. Peng, Anisotropic Strain Relaxation of Graphene by Corrugation on Copper Crystal Surfaces, Small 14(22) (2018) 1800725.

[53] U. Tartaglino, T. Zykova-Timan, F. Ercolessi, E. Tosatti, Melting and nonmelting of solid surfaces and nanosystems, Phys. Rep. 411(5) (2005) 291-321.

[54] A.D. Laine, M. DeSeta, C. Cepek, S. Vandré, A. Goldoni, N. Franco, J. Avila, M.C. Asensio, M. Sancrotti, Surface phase transitions of Ge(100) from temperature-dependent valence-band photoemission, Phys. Rev. B 57(23) (1998) 14654-14657.

[55] B. Kiraly, A.J. Mannix, R.M. Jacobberger, B.L. Fisher, M.S. Arnold, M.C. Hersam, N.P. Guisinger, Sub-5 nm, globally aligned graphene nanoribbons on Ge(001), Appl. Phys. Lett. 108(21) (2016) 213101.

[56] L. Persichetti, A. Sgarlata, M. Fanfoni, A. Balzarotti, Irreversible order-disorder transformation of Ge(0 0 1) probed by scanning tunnelling microscopy, J. Phys.: Condens. Matter 27(43) (2015) 435001.

[57] H.-B. Li, A.J. Page, C. Hettich, B. Aradi, C. Köhler, T. Frauenheim, S. Irle, K. Morokuma, Graphene nucleation on a surface-molten copper catalyst: quantum chemical molecular dynamics simulations, Chemical Science 5(9) (2014) 3493-3500.

[58] P.C. Rogge, K. Thürmer, M.E. Foster, K.F. McCarty, O.D. Dubon, N.C. Bartelt, Real-time observation of epitaxial graphene domain reorientation, Nature Communications 6 (2015) 6880.

[59] T. Niu, M. Zhou, J. Zhang, Y. Feng, W. Chen, Growth intermediates for CVD graphene on Cu(111): carbon clusters and defective graphene, J. Am. Chem. Soc. 135(22) (2013) 8409-8414.